\documentclass[conference]{IEEEtran}
\IEEEoverridecommandlockouts
\usepackage{cite}
\usepackage{amsmath,amssymb,amsfonts}
\usepackage{bm}
\usepackage{mathtools}
\usepackage{algorithmic}
\usepackage{graphicx}
\usepackage{textcomp}
\usepackage{float}
\usepackage{placeins}
\usepackage{multirow}
\usepackage{multicol}
\usepackage{booktabs}
\usepackage{xcolor}
\def\BibTeX{{\rm B\kern-.05em{\sc i\kern-.025em b}\kern-.08em
    T\kern-.1667em\lower.7ex\hbox{E}\kern-.125emX}}
\def\thline{\noalign{\hrule height 1.2pt}}
\begin{document}

\title{
    Audio Spotforming via Post-Filtering\\Using Cross-Array Non-target Estimates
    \thanks{
    Corresponding author: Yuto Ishikawa (yuto\_ishikawa.jp@ieee.org). This work was conducted during Yuto's internship at CyberAgent.
    }
}

\author{
    \IEEEauthorblockN{
        Yuto Ishikawa\IEEEauthorrefmark{1}\IEEEauthorrefmark{2},
        Li Li\IEEEauthorrefmark{1},
        Shogo Seki\IEEEauthorrefmark{1},
        and Kouei Yamaoka\IEEEauthorrefmark{2}
    }
    \IEEEauthorblockA{
        \IEEEauthorrefmark{1}CyberAgent, Inc., Japan
    }
    \IEEEauthorblockA{
        \IEEEauthorrefmark{2}The University of Tokyo, Japan
    }
}

\newcommand{\IndFreqbin}{i}
\newcommand{\NumFreqbins}{I}
\newcommand{\IndFrame}{j}
\newcommand{\NumFrames}{J}
\newcommand{\IndMic}{m}
\newcommand{\NumMics}{M}
\newcommand{\IndSrc}{n}
\newcommand{\NumSrcs}{N}
\newcommand{\IndBasis}{k}
\newcommand{\NumBases}{K}
\newcommand{\IndArray}{b}
\newcommand{\NumArrays}{B}

\newcommand{\IndBasisTarget}{k}
\newcommand{\NumBasesTarget}{K}
\newcommand{\IndBasisNoise}{l}
\newcommand{\NumBasesNoise}{L}

\newcommand{\MixMat}{\bm{A}}
\newcommand{\MixVec}{\bm{a}}
\newcommand{\DemixMat}{\bm{W}}
\newcommand{\DemixVec}{\bm{w}}
\newcommand{\Identity}{\bm{E}}
\newcommand{\UnitVec}{\bm{e}}

\newcommand{\atf}{\bm{h}}

\newcommand{\upt}{(\mathrm{t})}
\newcommand{\upn}{(\mathrm{n})}
\newcommand{\upx}{(\mathrm{x})}

\newcommand{\Hermite}{\mathsf{H}}
\newcommand{\Transpose}{\mathsf{T}}

\newcommand{\ObsSignal}{x}
\newcommand{\SepSignal}{y}
\newcommand{\SrcSignal}{s}
\newcommand{\TargetSignal}{s}
\newcommand{\NoiseSignal}{n}

\newcommand{\var}{\sigma}

\newcommand{\NMFBasis}{t}
\newcommand{\NMFActiv}{v}
\newcommand{\SrcModel}{r}

\newcommand{\Cost}[1]{\mathcal{L}_{#1}}

\newcommand{\Weight}{\lambda}
\newcommand{\ShapeParam}{\alpha}
\newcommand{\ScaleParam}{\beta}

\newcommand{\CovMat}{\bm{R}}
\newcommand{\Diag}{\bm{D}}
\newcommand{\IndTargetMic}{\IndMic^{\upt}}

\newcommand{\AuxJensenTarget}{\theta}
\newcommand{\AuxJensenNoise}{\phi}
\newcommand{\AuxLogX}{\eta}
\newcommand{\AuxLogN}{\zeta}
\newcommand{\AuxLogT}{\psi}

\allowdisplaybreaks

\maketitle

\begin{abstract}
Audio spotforming is a technique for extracting target speech from noisy mixtures by utilizing multiple microphone arrays.
Conventional methods estimate a shared target speech component from linearly separated signals obtained by each array using low-rank approximations and apply post filtering (PF) based on this estimated low-rank representation.
However, owing to the mismatch between low-rank models and the complex structure of speech signals, directly relying on low-rank approximations for PF can degrade the speech extraction performance.
In this study, we leverage the observation that non-target components located in the target speech direction from the perspective of one array can be spatially separated when viewed from other arrays. This insight motivates a new spotforming method for efficient post-filter estimation using non-target estimates across arrays instead of relying on low-rank approximations.
Experiments demonstrate that the proposed method outperforms conventional spotforming methods.
\end{abstract}


\vspace{-0.25em}
\section{Introduction}
\label{sec:introduction}

Multichannel target speech extraction (TSE) is a technique designed to isolate a target speech signal from multichannel recordings that contain interfering speech signals or background noise~\cite{2009TakahashiIEEE-TASLP}.
TSE has a wide range of applications, such as the preprocessing of hearable devices and automatic speech recognition in robot dialogue systems.
As a representative TSE approach, beamforming \cite{van2004optimum} using a single microphone array has been studied extensively for decades and is widely employed.
Beamforming forms spatial directivity toward the target speaker by exploiting the phase differences among microphones.
However, when non-target components (e.g., interfering speakers) are located in the same direction as the target speaker relative to the microphone array, beamforming struggles to spatially separate the target and non-target components~\cite{2003ArakiEURASIP-JASP}. Consequently, the TSE performance of the beamforming can be significantly degraded.
To address this limitation, recent studies introduced methods to enhance a specific spatial region~\cite{zheng2004robust,martinez2015robust,taseska2016spotforming,2022KagimotoIROS,2024AyanoEUSIPCO,gu2024rezero}.

One approach, \textit{audio spotforming}, leverages the ability of multiple distributed microphone arrays to capture different spatial characteristics. 
For instance, \cite{2022KagimotoIROS}  and \cite{2024AyanoEUSIPCO} utilize the presence of a target speech component across all the microphone arrays and employ a two-stage framework. 
This involves a spatial filtering (SF) stage to enhance the signals arriving from the target speech direction at each microphone array, and a post filtering (PF) stage to extract the target speech component shared across the microphone arrays from the outputs of the SF stage.
In \cite{2022KagimotoIROS}, nonnegative matrix factorization (NMF)~\cite{1999LeeNature}, a well-known low-rank approximation method, is incorporated into the PF stage to estimate the target speech component, based on the assumption that the spectral patterns of the target speech co-occur across multiple arrays.
To enhance model interpretability and robustness against signal misalignment among microphone arrays, \cite{2024AyanoEUSIPCO} employs nonnegative tensor factorization (NTF)~\cite{2007CichockiICANNGA} to represent the target and non-target components as distinct spectral patterns.
This relaxes the co-occurrence assumption and achieves high TSE performance.
These conventional methods assume a low-rank structure for the target speech signal and directly use the estimated low-rank representation to construct the post filters.
However, it is difficult to accurately represent the complex structure of speech signals using low-rank approximations~\cite{2016KitamuraIEEE-ACM-TASLP}.
In practice, \cite{2024AyanoEUSIPCO} reported that relaxing the low-rank approximation is necessary to achieve high performance, which increases the computational complexity and may reduce the efficiency in practical implementations.

\begin{figure}[t]
    \centering
    \includegraphics[width=0.8\columnwidth]{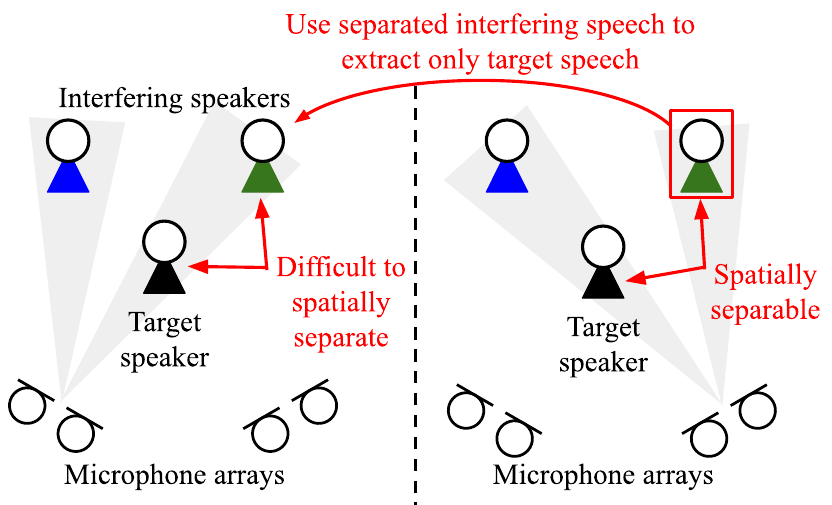}
    \vspace{-0.6em}
    \caption{
        Key concept of the proposed method.
        From the left microphone array, it is difficult to spatially separate the black target speaker and the green interfering speaker arriving from the same direction.
        By contrast, from the right microphone array, these two speakers are spatially separable.
        The proposed method exploits this property to extract only the target speech.
    }
    \vspace{-1.3em}
    \label{fig:key_concept}
\end{figure}

To achieve high TSE performance while reducing computational complexity, we propose a new spotforming method for efficient post-filter estimation that uses non-target components across microphone arrays.
Fig.~\ref{fig:key_concept} illustrates the key concept of the proposed method.
In scenarios where a method based on linear demixing filters is applied and an interfering speaker or diffuse noise is present in the same direction as the target speaker from the perspective of one array, the spatial separation of the target and non-target components is fundamentally challenging.
However, these components can be spatially separated from other microphone arrays.
The proposed method exploits this property to estimate both the target and non-target components arriving from the target speech direction without relying on a low-rank structure. 
Finally, a multichannel Wiener filter is constructed and applied to the observed signals at each microphone array.
Experiments demonstrated that the proposed method can extract a target speech signal more accurately than conventional spotforming methods.

\section{Related work: Audio Spotforming~\cite{2022KagimotoIROS,2024AyanoEUSIPCO}}

Let $\bm{\ObsSignal}_{\IndArray\IndFreqbin\IndFrame} = (\ObsSignal_{\IndArray\IndFreqbin\IndFrame1}, ..., \ObsSignal_{\IndArray\IndFreqbin\IndFrame\NumMics})^{\Transpose} \in \mathbb{C}^{\NumMics}$ be the short-time Fourier transform (STFT) coefficients of an observed signal.
Here, $\IndArray \in \{1, ..., \NumArrays \}$, $\IndFreqbin \in \{1, ..., \NumFreqbins\}$, $\IndFrame \in \{1, ..., \NumFrames\}$, and $\IndMic \in \{1, ..., \NumMics\}$ are the indices of the microphone arrays, frequency bins, time frames, and microphones, respectively, and $^{\Transpose}$ denotes the transpose.
In the situation where the non-target components (e.g., interfering speakers and diffuse noise) exist around the target speaker, the observed signal at each microphone array can be approximately modeled as $\bm{\ObsSignal}_{\IndArray\IndFreqbin\IndFrame} = \bm{\TargetSignal}_{\IndArray\IndFreqbin\IndFrame} + \bm{\NoiseSignal}_{\IndArray\IndFreqbin\IndFrame}$, where $\bm{\TargetSignal}_{\IndArray\IndFreqbin\IndFrame} \in \mathbb{C}^{\NumMics}$ denotes the source image of the target speech indicated by positional information, and $\bm{\NoiseSignal}_{\IndArray\IndFreqbin\IndFrame} \in \mathbb{C}^{\NumMics}$ represents the sum of the source images of all remaining non-target components.
In particular, if a point-source interfering component is located in the same direction as the target speaker from the perspective of a microphone array, it is fundamentally difficult to spatially separate the target and non-target components~\cite{2003ArakiEURASIP-JASP}.
To address this limitation, spotforming methods~\cite{2022KagimotoIROS,2024AyanoEUSIPCO} extract the target speech signal in two stages: the SF and PF stages.
In the SF stage, a spatial filter is applied to the observed signals at each microphone array to enhance signals arriving from the target speech direction using beamforming or methods based on linear demixing filters.
In the PF stage, a post filter is applied to the outputs of the SF stage to further suppress the residual non-target components and enhance the target speech components shared across arrays.
Conventional methods~\cite{2022KagimotoIROS,2024AyanoEUSIPCO} use low-rank approximations to estimate the post filter.
Specifically, \cite{2022KagimotoIROS} employs NMF, whereas \cite{2024AyanoEUSIPCO} uses NTF to model the common spectral structure of the target speech. 
A Wiener filter is then constructed based on the estimated low-rank representation.
Finally, the extracted signal is obtained by synchronously averaging the outputs of the PF stage.

\section{Proposed method}
\label{sec:proposed_method}

\subsection{Motivations}
\label{ssec:motivations}

Conventional spotforming methods~\cite{2022KagimotoIROS,2024AyanoEUSIPCO} successfully model the spectral structures of the target speech shared across arrays using low-rank approximations. 
However, their performance highly depends on the model complexity, particularly the selection of the number of bases, owing to the direct use of the estimated low-rank representation to construct post filters.
This dependency was experimentally confirmed in \cite{2024AyanoEUSIPCO}, which reported that a large number of bases were necessary to achieve adequate modeling accuracy.
Instead of relying on low-rank approximations, we exploit the observation that when a method based on linear demixing filters is applied to the mixtures observed at each array, non-target components arriving from the target speech direction in one array can be spatially separated in other arrays.
This insight motivates a new spotforming method that utilizes cross-array non-target estimates rather than low-rank models.

\subsection{Requirements for SF stage}
\label{ssec:requirement_bf_part}

Unlike conventional spotforming methods, which utilize only spatially separated signals corresponding to the target speech in the PF stage, the proposed method additionally requires non-target components estimated in the SF stage.
For this purpose, we employ a method based on linear demixing filters, such as independent low-rank matrix analysis (ILRMA)~\cite{2016KitamuraIEEE-ACM-TASLP}, rather than a conventional beamformer.

Let us consider time-invariant methods,
where two types of spatial filters are obtained: a spatial filter $\DemixVec_{\IndArray\IndFreqbin\IndTargetMic} \in \mathbb{C}^{\NumMics}$ derived to enhance the source(s) located in the target speech direction, and $\NumMics - 1$ spatial filters $\DemixVec_{\IndArray\IndFreqbin\IndMic} \in \mathbb{C}^{\NumMics}\ (\IndMic \ne \IndTargetMic)$ aimed at suppressing the signal at the target speech direction while enhancing the dominant non-target components.
Here, $\IndTargetMic$ denotes the channel index corresponding to the target speech direction.
For convenience, we represent the set of spatial filters at the $\IndArray$th array as $\DemixMat_{\IndArray\IndFreqbin} \coloneq (\DemixVec_{\IndArray\IndFreqbin1}, ..., \DemixVec_{\IndArray\IndFreqbin\NumMics})^{\Hermite} \in \mathbb{C}^{\NumMics \times \NumMics}$, where $^{\Hermite}$ denotes the Hermitian transpose.
To ensure that a specific spatial filter consistently corresponds to the target speech direction, we employ spatially regularized linear demixing filter-based methods. 
Specifically, we adopt the geometrically constrained ILRMA (GC-ILRMA)~\cite{2018MitsuiICASSP}, which is an ILRMA variant regularized by a direction-based spatial prior.

\subsection{Formulation in PF stage}
\label{ssec:formulation}

We assume that the observed signal follows a zero-mean multivariate complex Gaussian distribution as 
\begin{align}
    \bm{\ObsSignal}_{\IndArray\IndFreqbin\IndFrame} \sim \mathcal{N}_{\mathrm{MC}} \bigl( \bm{0}_{\NumMics}, \CovMat_{\IndArray\IndFreqbin\IndFrame}^{\upx} \bigr),
\end{align}
where $\CovMat_{bij}^{\upx} \in \mathbb{C}^{\NumMics \times \NumMics}$ is the covariance matrix of the observed signal.
To represent the non-target component arriving from the target speech direction, we model $\CovMat_{\IndArray\IndFreqbin\IndFrame}^{\upx}$ as follows:
\begin{align}
    \label{eq:def:CovMat_Obs}
    \CovMat_{\IndArray\IndFreqbin\IndFrame}^{\upx} &= \MixMat_{\IndArray\IndFreqbin} \Diag_{\IndArray\IndFreqbin\IndFrame} \MixMat_{\IndArray\IndFreqbin}^{\Hermite},
    \\
    \label{eq:def:DiagMat}
    \Diag_{\IndArray\IndFreqbin\IndFrame} &= \SrcModel_{\IndFreqbin\IndFrame}^{\upt} \UnitVec_{\IndTargetMic} \UnitVec_{\IndTargetMic}^{\Hermite} + \mathrm{diag}\bigl( \SrcModel_{\IndArray\IndFreqbin\IndFrame1}^{\upn}, ..., \SrcModel_{\IndArray\IndFreqbin\IndFrame\NumMics}^{\upn} \bigr), 
\end{align}
where $\UnitVec_{\IndMic} \in \mathbb{R}^{\NumMics}$ is a one-hot vector whose $\IndMic$th element is one and the others are zero, $\MixMat_{\IndArray\IndFreqbin} \coloneq \DemixMat_{\IndArray\IndFreqbin}^{-1} \in \mathbb{C}^{\NumMics \times \NumMics}$ represents the time-invariant spatial characteristics from the perspective of the $\IndArray$th microphone array estimated in the SF stage, and $\SrcModel_{\IndFreqbin\IndFrame}^{\upt} > 0$ and $\SrcModel_{\IndArray\IndFreqbin\IndFrame\IndMic}^{\upn} > 0$ denote the time-varying variances of the target speech and non-target components, respectively.
Here, $\SrcModel_{\IndArray\IndFreqbin\IndFrame\IndTargetMic}^{\upn}$ corresponds to the non-target component located in the target speech direction.
To distinguish the target speech from non-target components arriving in the same direction, we model $\SrcModel_{\IndArray\IndFreqbin\IndFrame\IndTargetMic}^{\upn}$ using the non-target components from the perspective of the other $\NumArrays - 1$ microphone arrays, as
\begin{align}
    \label{eq:SrcModel_Noise_at_TargetMicIndex}
    \SrcModel_{\IndArray\IndFreqbin\IndFrame\IndTargetMic}^{\upn} &= \sum_{\IndArray' \ne \IndArray, \IndMic \ne \IndTargetMic} \Weight_{\IndArray\IndArray'\IndMic} \SrcModel_{\IndArray'\IndFreqbin\IndFrame\IndMic}^{\upn},
\end{align}
where $\Weight_{\IndArray\IndArray'\IndMic} \geq 0$ is a scalar variable.
As in \cite{2020KuboIEEE-ACM-TASLP}, we further introduce the inverse gamma distribution as a prior distribution for $\SrcModel_{\IndFreqbin\IndFrame}^{\upt}$  to induce sparsity of the target speech, given as
\begin{align}
    \label{eq:prior_dist:SrcModel_target}
    \SrcModel_{\IndFreqbin\IndFrame}^{\upt} \sim \mathcal{IG}(\ShapeParam, \ScaleParam),
\end{align}
where $\ShapeParam > 0$ and $\ScaleParam > 0$ are the shape and scale parameters of the inverse gamma distribution, respectively.

The cost function $\Cost{}$ is defined as the negative log-posterior of the observed signal over all microphone arrays with the prior distribution of $\SrcModel_{\IndFreqbin\IndFrame}^{\upt}$:
\begin{align}
    \label{eq:Cost:Proposed}
    \Cost{} &= \sum_{\IndArray,\IndFreqbin,\IndFrame} \Bigl[
        \bm{\ObsSignal}_{\IndArray\IndFreqbin\IndFrame}^{\Hermite} \bigl( \CovMat_{\IndArray\IndFreqbin\IndFrame}^{\upx} \bigr)^{-1} \bm{\ObsSignal}_{\IndArray\IndFreqbin\IndFrame} + \log{\det{\CovMat_{\IndArray\IndFreqbin\IndFrame}^{\upx}}}
    \Bigr]
    \nonumber\\
    &\phantom{=} + \sum_{\IndFreqbin,\IndFrame} \Biggl[ (\ShapeParam + 1) \log{\SrcModel_{\IndFreqbin\IndFrame}^{\upt}} + \frac{\ScaleParam}{\SrcModel_{\IndFreqbin\IndFrame}^{\upt}} \Biggr].
\end{align}
Here, $\SrcModel_{\IndFreqbin\IndFrame}^{\upt}$, $\SrcModel_{\IndArray\IndFreqbin\IndFrame\IndMic}^{\upn}$, and $\Weight_{\IndArray\IndArray'\IndMic}$ $(\IndArray' \ne \IndArray, \IndMic \ne  \IndTargetMic)$ are objective variables to be estimated. Note that, for simplicity, we exclude terms independent of the objective variables. 
By defining the spatially separated signal estimated in the SF stage $\bm{\SepSignal}_{\IndArray\IndFreqbin\IndFrame} = (\SepSignal_{\IndArray\IndFreqbin\IndFrame1}, ..., \SepSignal_{\IndArray\IndFreqbin\IndFrame\NumMics})^{\Transpose} \in \mathbb{C}^{\NumMics}$ as 
\begin{align}
    \label{eq:def:SepSignal}
    \bm{\SepSignal}_{\IndArray\IndFreqbin\IndFrame} &= \DemixMat_{\IndArray\IndFreqbin} \bm{\ObsSignal}_{\IndArray\IndFreqbin\IndFrame},
\end{align}
and substituting (\ref{eq:def:CovMat_Obs})--(\ref{eq:SrcModel_Noise_at_TargetMicIndex}) and (\ref{eq:def:SepSignal}) into (\ref{eq:Cost:Proposed}), $\Cost{}$ can be rewritten as
\begin{align}
    \label{eq:Cost:Proposed_transform}
    &\Cost{} = \sum_{\IndArray,\IndFreqbin,\IndFrame} \Biggl[
        \frac{| \SepSignal_{\IndArray\IndFreqbin\IndFrame\IndTargetMic} |^{2}}{\SrcModel_{\IndFreqbin\IndFrame}^{\upt} + \sum_{\IndArray' \ne \IndArray, \IndMic' \ne \IndTargetMic} \Weight_{\IndArray\IndArray'\IndMic'} \SrcModel_{\IndArray'\IndFreqbin\IndFrame\IndMic'}^{\upn}}
    \nonumber\\
        &\phantom{} + \sum_{\IndMic \ne \IndTargetMic} \frac{| \SepSignal_{\IndArray\IndFreqbin\IndFrame\IndMic} |^{2}}{\SrcModel_{\IndArray\IndFreqbin\IndFrame\IndMic}^{\upn}} + \log{\biggl( \SrcModel_{\IndFreqbin\IndFrame}^{\upt} + \sum_{\IndArray' \ne \IndArray, \IndMic' \ne \IndTargetMic} \Weight_{\IndArray\IndArray'\IndMic'} \SrcModel_{\IndArray'\IndFreqbin\IndFrame\IndMic'}^{\upn} \biggr)}
    \nonumber\\
    &\phantom{} + \sum_{\IndMic \ne \IndTargetMic} \log{\SrcModel_{\IndArray\IndFreqbin\IndFrame\IndMic}^{\upn}} + \log{ \det{\bigl( \MixMat_{\IndArray\IndFreqbin} \MixMat_{\IndArray\IndFreqbin}^{\Hermite} \bigr)}}
    \Biggr]
    \nonumber\\
    &\phantom{} + \sum_{\IndFreqbin,\IndFrame} \Biggl[ 
        (\ShapeParam + 1) \log{\SrcModel_{\IndFreqbin\IndFrame}^{\upt}} + \frac{\ScaleParam}{\SrcModel_{\IndFreqbin\IndFrame}^{\upt}}
    \Biggr].
\end{align}

After estimating the parameters, we extract the source image of the target speech signal using a multichannel Wiener filter: 
\begin{align}
    \label{eq:MWF:proposed}
    \hat{\bm{\SrcSignal}}_{\IndArray\IndFreqbin\IndFrame} &= \SrcModel_{\IndFreqbin\IndFrame}^{\upt} \MixMat_{\IndArray\IndFreqbin} \UnitVec_{\IndTargetMic} \UnitVec_{\IndTargetMic}^{\Hermite} \MixMat_{\IndArray\IndFreqbin}^{\Hermite} \bigl( \CovMat_{\IndArray\IndFreqbin\IndFrame}^{\upx} \bigr)^{-1} \bm{\ObsSignal}_{\IndArray\IndFreqbin\IndFrame}
    \nonumber\\
    &= \frac{\SrcModel_{\IndFreqbin\IndFrame}^{\upt}}{\SrcModel_{\IndFreqbin\IndFrame}^{\upt} + \sum_{\IndArray' \ne \IndArray, \IndMic \ne \IndTargetMic} \Weight_{\IndArray\IndArray'\IndMic} \SrcModel_{\IndArray'\IndFreqbin\IndFrame\IndMic}^{\upn}} \bm{\SepSignal}_{\IndArray\IndFreqbin\IndFrame}.
\end{align}
The final output is obtained by synchronously averaging the extracted signals across all microphone arrays in the time domain.

\subsection{Derivation of update rules for iterative algorithm}
\label{ssec:derivation_of_update_rule}

Since it is difficult to directly minimize the cost function, we employ the majorization-equalization (ME) algorithm~\cite{2011FevotteNC} and estimate the objective variables in an iterative manner.
By utilizing Jensen's inequality and the relationship between a concave function and its tangent line, $\log{x} \leq (x - c) / c + \log{c}\ (\forall x, c \in \mathbb{R}_{\geq 0})$, we design the following auxiliary function of $\Cost{}$:
\begin{align}
    \label{eq:AuxCost:Proposed}
    &\bar{\Cost{}} = 
    \sum_{\IndArray,\IndFreqbin,\IndFrame} \Biggl[
        \frac{| \SepSignal_{\IndArray\IndFreqbin\IndFrame\IndTargetMic} |^{2} \AuxJensenTarget_{\IndArray\IndFreqbin\IndFrame}^{2}}{\SrcModel_{\IndFreqbin\IndFrame}^{\upt}} 
    \nonumber\\
        &\phantom{} + \hspace{-0.2em} \sum_{\IndArray' \ne \IndArray, \IndMic' \ne \IndTargetMic} \hspace{-0.5em} \frac{
            | \SepSignal_{\IndArray\IndFreqbin\IndFrame\IndTargetMic} |^{2} (1 - \AuxJensenTarget_{\IndArray\IndFreqbin\IndFrame})^{2} \AuxJensenNoise_{\IndArray\IndArray'\IndFreqbin\IndFrame\IndMic'}^{2}
        }{
            \Weight_{\IndArray\IndArray'\IndMic'} \SrcModel_{\IndArray'\IndFreqbin\IndFrame\IndMic'}^{\upn}
        } \hspace{-0.1em} + \hspace{-0.3em} \sum_{\IndMic \ne \IndTargetMic} \hspace{-0.3em} \frac{| \SepSignal_{\IndArray\IndFreqbin\IndFrame\IndMic} |^{2}}{\SrcModel_{\IndArray\IndFreqbin\IndFrame\IndMic}^{\upn}}
    \nonumber\\
    &\phantom{} + \frac{
        \SrcModel_{\IndFreqbin\IndFrame}^{\upt} + \sum_{\IndArray' \ne \IndArray, \IndMic' \ne \IndTargetMic} \Weight_{\IndArray\IndArray'\IndMic'} \SrcModel_{\IndArray'\IndFreqbin\IndFrame\IndMic'}^{\upn} - \AuxLogX_{\IndArray\IndFreqbin\IndFrame}
    }{\AuxLogX_{\IndArray\IndFreqbin\IndFrame}} + \log{\AuxLogX_{\IndArray\IndFreqbin\IndFrame}}
    \nonumber\\
    &\phantom{} + \hspace{-0.2em} \sum_{\IndMic \ne \IndTargetMic} \Biggl(
        \frac{
            \SrcModel_{\IndArray\IndFreqbin\IndFrame\IndMic}^{\upn} - \AuxLogN_{\IndArray\IndFreqbin\IndFrame\IndMic}
        }{
            \AuxLogN_{\IndArray\IndFreqbin\IndFrame\IndMic}
        } + \log{\AuxLogN_{\IndArray\IndFreqbin\IndFrame\IndMic}}
    \Biggr) + \log{\det{\bigl( \MixMat_{\IndArray\IndFreqbin} \MixMat_{\IndArray\IndFreqbin}^{\Hermite} \bigr)}}
    \Biggr]
    \nonumber\\
    &\phantom{} + \sum_{\IndFreqbin,\IndFrame} \Biggl[
        (\ShapeParam + 1) \Biggl(
            \frac{\SrcModel_{\IndFreqbin\IndFrame}^{\upt} - \AuxLogT_{\IndFreqbin\IndFrame}}{\AuxLogT_{\IndFreqbin\IndFrame}} + \log{\AuxLogT_{\IndFreqbin\IndFrame}}
        \Biggr) + \frac{\ScaleParam}{\SrcModel_{\IndFreqbin\IndFrame}^{\upt}}
    \Biggr],
\end{align}
where $\AuxJensenTarget_{\IndArray\IndFreqbin\IndFrame} \in [0, 1]$, $\AuxJensenNoise_{\IndArray\IndArray'\IndFreqbin\IndFrame\IndMic} \in [0, 1]$, $\AuxLogX_{\IndArray\IndFreqbin\IndFrame} > 0$, $\AuxLogN_{\IndArray\IndFreqbin\IndFrame\IndMic} \allowbreak > \allowbreak 0$, and $\AuxLogT_{\IndFreqbin\IndFrame} \allowbreak > \allowbreak 0$ are auxiliary variables satisfying $\sum_{\IndArray' \ne \IndArray, \IndMic \ne \IndTargetMic} \AuxJensenNoise_{\IndArray\IndArray'\IndFreqbin\IndFrame\IndMic} = 1$.
The equality $\bar{\Cost{}} = \Cost{}$ holds if and only if the following conditions hold:
\begin{align}
    \label{eq:AuxJensenTarget:Condition}
    \AuxJensenTarget_{\IndArray\IndFreqbin\IndFrame} &= \frac{\SrcModel_{\IndFreqbin\IndFrame}^{\upt}}{\SrcModel_{\IndFreqbin\IndFrame}^{\upt} + \sum_{\IndArray' \ne \IndArray, \IndMic' \ne \IndTargetMic} \Weight_{\IndArray\IndArray'\IndMic'} \SrcModel_{\IndArray'\IndFreqbin\IndFrame\IndMic'}^{\upn}},
    \\
    \label{eq:AuxJensenNoise:Condition}
    \AuxJensenNoise_{\IndArray\IndArray'\IndFreqbin\IndFrame\IndMic} &= \frac{
        \Weight_{\IndArray\IndArray'\IndMic} \SrcModel_{\IndArray'\IndFreqbin\IndFrame\IndMic}^{\upn}
    }{
        \sum_{\IndArray'' \ne \IndArray, \IndMic' \ne \IndTargetMic} \Weight_{\IndArray\IndArray''\IndMic'} \SrcModel_{\IndArray''\IndFreqbin\IndFrame\IndMic'}^{\upn}
    },
    \\
    \label{eq:AuxLogX:Condition}
    \AuxLogX_{\IndArray\IndFreqbin\IndFrame} &= \SrcModel_{\IndFreqbin\IndFrame}^{\upt} + \sum_{\IndArray' \ne \IndArray, \IndMic' \ne \IndTargetMic} \Weight_{\IndArray\IndArray'\IndMic'} \SrcModel_{\IndArray'\IndFreqbin\IndFrame\IndMic'}^{\upn},
    \\
    \label{eq:AuxLogN:Condition}
    \AuxLogN_{\IndArray\IndFreqbin\IndFrame\IndMic} &= \SrcModel_{\IndArray\IndFreqbin\IndFrame\IndMic}^{\upn},
    \\
    \label{eq:AuxLogT:Condition}
    \AuxLogT_{\IndFreqbin\IndFrame} &= \SrcModel_{\IndFreqbin\IndFrame}^{\upt}.
\end{align}
Here, the auxiliary function $\bar{\Cost{}}$ consists of linear and reciprocal terms with respect to each objective variable, and has the same form as the auxiliary function in \cite{2020KuboIEEE-ACM-TASLP}.
Therefore, the update rules based on the ME algorithm can be derived in the same manner as in \cite{2020KuboIEEE-ACM-TASLP}.
Using (\ref{eq:AuxJensenTarget:Condition})--(\ref{eq:AuxLogT:Condition}), we finally obtain the following update rules:
\begin{align}
    \label{eq:UpdateRule:SrcModel_Target:Proposed}
    \SrcModel_{\IndFreqbin\IndFrame}^{\upt} &\leftarrow \SrcModel_{\IndFreqbin\IndFrame}^{\upt} \Biggl( \frac{
        \sum_{\IndArray} | \SepSignal_{\IndArray\IndFreqbin\IndFrame\IndTargetMic} |^{2} \tilde{\SrcModel}_{\IndArray\IndFreqbin\IndFrame}^{-2} + \ScaleParam \bigl( \SrcModel_{\IndFreqbin\IndFrame}^{\upt} \bigr)^{-2}
    }{
        \sum_{\IndArray} \tilde{\SrcModel}_{\IndArray\IndFreqbin\IndFrame}^{-1} + (\ShapeParam + 1) \bigl( \SrcModel_{\IndFreqbin\IndFrame}^{\upt} \bigr)^{-1}
    } 
    \Biggr),
    \\
    \label{eq:UpdateRule:SrcModel_Noise:Proposed}
    \SrcModel_{\IndArray\IndFreqbin\IndFrame\IndMic}^{\upn} &\leftarrow \SrcModel_{\IndArray\IndFreqbin\IndFrame\IndMic}^{\upn}
    \nonumber\\
    &\phantom{}\cdot \Biggl( \frac{
        \sum_{\IndArray' \ne \IndArray} | \SepSignal_{\IndArray'\IndFreqbin\IndFrame\IndTargetMic} |^{2} \Weight_{\IndArray'\IndArray\IndMic} \tilde{\SrcModel}_{\IndArray'\IndFreqbin\IndFrame}^{-2} + | \SepSignal_{\IndArray\IndFreqbin\IndFrame\IndMic} |^{2} \bigl( \SrcModel_{\IndArray\IndFreqbin\IndFrame\IndMic}^{\upn} \bigr)^{-2}
    }{
        \sum_{\IndArray' \ne \IndArray} \Weight_{\IndArray'\IndArray\IndMic} \tilde{\SrcModel}_{\IndArray'\IndFreqbin\IndFrame}^{-1} + \bigl( \SrcModel_{\IndArray\IndFreqbin\IndFrame\IndMic}^{\upn} \bigr)^{-1}
    }
    \Biggr),
    \\
    \label{eq:UpdateRule:Weight:Proposed}
    \Weight_{\IndArray\IndArray'\IndMic} &\leftarrow \Weight_{\IndArray\IndArray'\IndMic} \Biggl( \frac{
        \sum_{\IndFreqbin,\IndFrame} | \SepSignal_{\IndArray\IndFreqbin\IndFrame\IndTargetMic} |^{2} \SrcModel_{\IndArray'\IndFreqbin\IndFrame\IndMic}^{\upn} \tilde{\SrcModel}_{\IndArray\IndFreqbin\IndFrame}^{-2}
    }{
        \sum_{\IndFreqbin,\IndFrame} \SrcModel_{\IndArray'\IndFreqbin\IndFrame\IndMic}^{\upn} \tilde{\SrcModel}_{\IndArray\IndFreqbin\IndFrame}^{-1}
    }
    \Biggr),
\end{align}
where $\tilde{\SrcModel}_{\IndArray\IndFreqbin\IndFrame}$ is an intermediate variable defined as
\vspace{-0.25em}
\begin{align}
    \label{eq:def:TildeSrcModel}
    \tilde{\SrcModel}_{\IndArray\IndFreqbin\IndFrame} &= \SrcModel_{\IndFreqbin\IndFrame}^{\upt} + \sum_{\IndArray' \ne \IndArray, \IndMic \ne \IndTargetMic} \Weight_{\IndArray\IndArray'\IndMic} \SrcModel_{\IndArray'\IndFreqbin\IndFrame\IndMic}^{\upn}
\end{align}
and is recalculated after each update of the objective variables.
For the initialization, we use the signal output at the SF stage as follows:
\vspace{-0.25em}
\begin{align}
    \label{eq:Initialization:SrcModel_Target}
    \SrcModel_{\IndFreqbin\IndFrame}^{\upt} &= \frac{1}{\NumArrays} \sum_{\IndArray} | \SepSignal_{\IndArray\IndFreqbin\IndFrame\IndTargetMic} |^{2},
    \\
    \SrcModel_{\IndArray\IndFreqbin\IndFrame\IndMic}^{\upn} &= | \SepSignal_{\IndArray\IndFreqbin\IndFrame\IndMic} |^{2}, \ (\IndMic \ne \IndTargetMic)
    \\
    \Weight_{\IndArray\IndArray'\IndMic} &= \begin{cases}
        (\NumArrays - 1)^{-1},      &\text{(if $\IndArray' \ne \IndArray$ and $\IndMic \ne \IndTargetMic$)}
        \\
        0.                      &\text{(otherwise)}
    \end{cases}
\end{align}

\section{Experiments}
\label{sec:experiment}

\vspace{-0.1em}
\subsection{Experimental settings}

\begin{figure}[t]
    \centering
    \includegraphics[width=0.75\columnwidth]{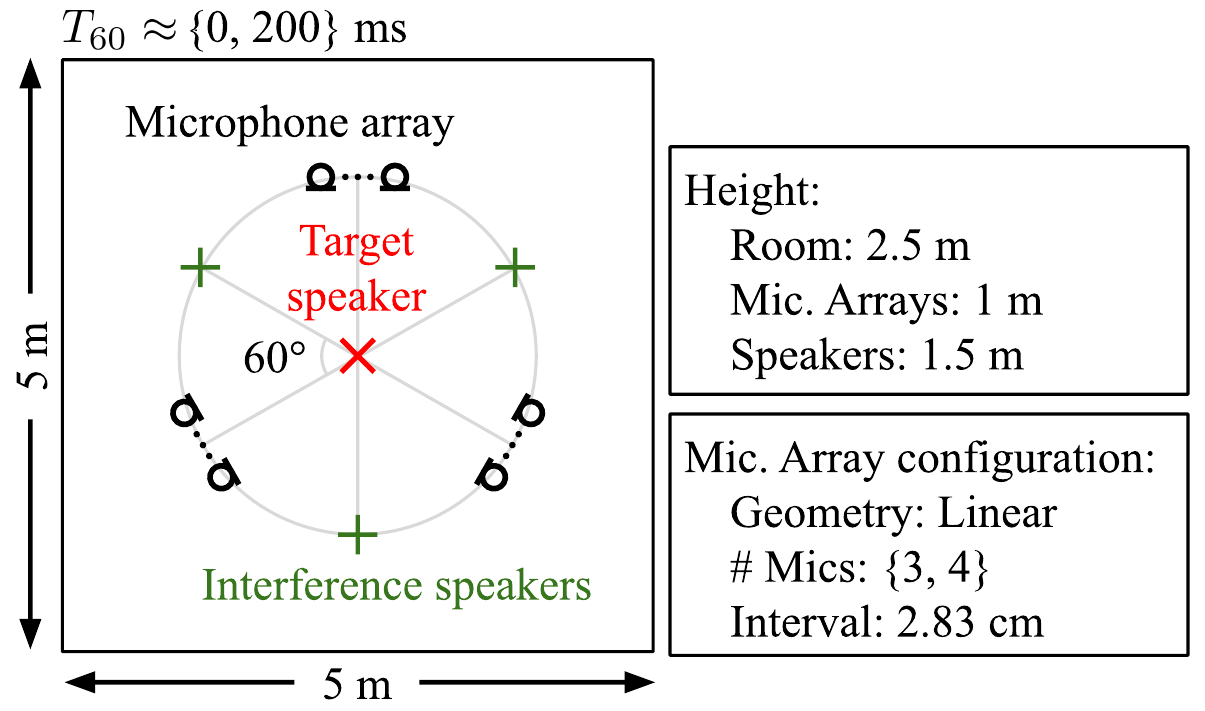}
    \vspace{-0.7em}
    \caption{
        Room layout for simulating impulse responses.
        }
    \label{fig:room_layout}
    \vspace{-1.5em}
\end{figure}

We conducted four experiments by combining the two reverberation conditions with two different numbers of microphones per array.
Speech signals from 100 speakers in the JVS dataset~\cite{2020TakamichiAST} were used in this study.
For each target speaker, three speakers were randomly selected as the interference speakers.
A 5-s-long dry-source signal was created for each speaker by concatenating multiple utterances.
All speech signals were convolved with room impulse responses generated using the image source method implemented in Pyroomacoustics~\cite{2018ScheiblerICASSP}.
We assume that the sampling time/rate offsets among the distributed microphone arrays have been roughly compensated in advance, and residual asynchrony due to synchronization errors remains small.
To simulate this residual asynchrony, a random temporal shift of up to 10 samples was applied to the source images at each array.
The target and non-target signals were then mixed with equal power, where the power was computed by averaging over all microphones across all arrays.
In total, 100 samples were created for each case.
Fig.~\ref{fig:room_layout} shows the room layout used in the simulations.
The reverberation time $T_{60}$ was set to approximately 0~ms for cases (i) and (iii) to verify the theoretical performance limitation, and to 200~ms for cases (ii) and (iv) to simulate realistic scenarios.
The number of microphone arrays was set to three, with three microphones per array in cases (i) and (ii), and four microphones per array in cases (iii) and (iv).
The microphones were arranged with a spacing of 2.83~cm.
The sampling rate was set to 16~kHz.
The STFT was performed using a 64-ms-long Hann window with a shift length of 32~ms. 

We compared three methods: 
\textit{Conv. (NMF)} and \textit{Conv. (NTF)}, which are conventional methods using NMF~\cite{2022KagimotoIROS} and NTF~\cite{2024AyanoEUSIPCO}, respectively, and \textit{Prop.}, which is the proposed method.
All the compared methods utilized GC-ILRMA in the SF stage.
The number of bases in GC-ILRMA was set to 10.
For the PF stage, the number of bases was set to 50 for Conv. (NMF) and 300 for Conv. (NTF) on the basis of the experimental results reported in \cite{2024AyanoEUSIPCO}.
The number of iterations was set to 100 for GC-ILRMA and for the PF stage of the conventional methods as in \cite{2024AyanoEUSIPCO}.
That for the PF stage of the proposed method was experimentally set to 20.
The target channel index $\IndTargetMic$ was set to $\NumMics$.
For GC-ILRMA in the SF stage, the prior steering vector of the target speech was computed under free-field propagation with distance-based attenuation, using the same horizontal configuration as that of the simulated target speaker.
For the inverse gamma distribution used in Prop., the scale parameter $\ScaleParam$ was fixed to $10^{-16}$, and the shape parameter $\ShapeParam$ was set to $1$, $10^{-1}$, $10^{-2}$, and $10^{-3}$.
In addition, the proposed method without a prior distribution was evaluated for ablation.
The basis and activation variables of the NMF and NTF were initialized with uniformly distributed random values in the range of $[10^{-10}, 1]$. 
To account for the initialization dependence of the NMF and NTF variables, we used 10 random seeds for each sample.
The spatial filter $\DemixMat_{\IndArray\IndFreqbin}$ was initialized as the inverse of a matrix consisting of the prior steering vector of the target speech and its orthogonal complements.

For evaluation metrics, we used the source-to-distortion ratio (SDR), source-to-interferences ratio (SIR)~\cite{2006VincentIEEE-TASLP}, perceptual evaluation of speech quality (PESQ)~\cite{rix2001perceptual}, and short-time objective intelligibility (STOI)~\cite{2011TaalIEEE-TASLP}.
These metrics were computed for each array using the corresponding reference signals and then averaged across all the arrays for each sample.

\vspace{-0.1em}
\subsection{Comparison with conventional methods}
\label{ssec:comparison_w_conventional_methods}

Tables~\ref{table:cases_1_2} and \ref{table:cases_3_4} summarize the evaluation scores for cases (i) and (ii), and cases (iii) and (iv), respectively.
The values are presented as the mean $\pm$ standard deviation across 1000 scores.
Here, ``w/o prior" refers to the variant of the proposed method without the prior distribution.
The highest scores for each metric are highlighted in bold.
In most cases, Prop. achieved higher performance than the conventional methods.
Specifically, Prop. (w/o prior) tended to achieve higher SDR, PESQ, and STOI scores, whereas Prop. with a prior distribution ($\ShapeParam = 1, ..., 10^{-3}$) yielded higher SIR scores.
By inducing sparsity in the time-varying variance of the target speech through the prior distribution, non-target components can be effectively suppressed.
However, the use of a larger shape parameter excessively enforces sparsity, which may also attenuate the target speech component itself and introduce distortion into the extracted signal. 
This trade-off explains the observed differences in the evaluation metrics.

\begin{table*}[t]
\centering
\caption{
    Results in cases (i) and (ii): 3 microphones per array, with $T_{60} \approx$ 0 and 200~ms, respectively.
}
\label{table:cases_1_2}
\vspace{-1em}
\begin{tabular}{clcccccccc}
\thline
\multicolumn{2}{c}{\multirow{2}{*}[-1ex]{Method}} & \multicolumn{4}{c}{\multirow{1}{*}[-0.5ex]{Case~(i)}} & \multicolumn{4}{c}{\multirow{1}{*}[-0.5ex]{Case~(ii)}}
\\
\cmidrule(lr){3-6}\cmidrule(lr){7-10}
\multicolumn{2}{c}{}    & SDR {[}dB{]} & SIR {[}dB{]} & PESQ & STOI {[}\%{]} & SDR {[}dB{]} & SIR {[}dB{]} & PESQ & STOI {[}\%{]}
\\
\midrule
\multicolumn{2}{c}{Input} 
& 0.13 $\pm$ 0.12 & 0.13 $\pm$ 0.12 & 1.10 $\pm$ 0.05 & 53.12 $\pm$ 6.21 
& 0.13 $\pm$ 0.13 & 0.13 $\pm$ 0.13 & 1.12 $\pm$ 0.05 & 51.01 $\pm$ 6.28 
\\
\multicolumn{2}{c}{Conv. (NMF)~\cite{2022KagimotoIROS}} 
& 15.79 $\pm$ 0.53 & 18.89 $\pm$ 0.78 & 1.96 $\pm$ 0.25 & 89.80 $\pm$ 3.13 
& 7.13 $\pm$ 1.42 & 13.19 $\pm$ 1.76 & 1.32 $\pm$ 0.14 & 64.01 $\pm$ 5.89 
\\
\multicolumn{2}{c}{Conv. (NTF)~\cite{2024AyanoEUSIPCO}} 
& 16.88 $\pm$ 1.78 & 29.79 $\pm$ 2.38 & 3.06 $\pm$ 0.28 & 94.74 $\pm$ 2.40 
& 7.47 $\pm$ 1.61 & 20.03 $\pm$ 2.03 & \textbf{1.44 $\pm$ 0.16} & 64.32 $\pm$ 6.23 
\\
\hline
\multirow{5}{*}{Prop.} 
& $\ShapeParam = 1$
& 15.73 $\pm$ 2.07 & 29.51 $\pm$ 2.62 & 2.21 $\pm$ 0.34 & 91.58 $\pm$ 2.49 
& 5.58 $\pm$ 1.75 & \textbf{20.54 $\pm$ 3.99} & 1.15 $\pm$ 0.06 & 58.27 $\pm$ 1.78 
\\
& $\ShapeParam = 10^{-1}$
& 18.42 $\pm$ 1.94 & 30.99 $\pm$ 2.57 & 2.62 $\pm$ 0.32 & 94.41 $\pm$ 1.84 
& 6.96 $\pm$ 1.82 & 19.56 $\pm$ 3.79 & 1.25 $\pm$ 0.11 & 63.44 $\pm$ 6.41 
\\
& $\ShapeParam = 10^{-2}$
& 18.71 $\pm$ 1.90 & \textbf{31.10 $\pm$ 2.54} & 2.67 $\pm$ 0.31 & 94.68 $\pm$ 1.77 
& 7.10 $\pm$ 1.82 & 19.42 $\pm$ 3.75 & 1.26 $\pm$ 0.11 & 63.96 $\pm$ 6.38 
\\
& $\ShapeParam = 10^{-3}$
& 18.75 $\pm$ 1.90 & \textbf{31.10 $\pm$ 2.54} & 2.67 $\pm$ 0.31 & 94.70 $\pm$ 1.77 
& 7.11 $\pm$ 1.82 & 19.40 $\pm$ 3.74 & 1.27 $\pm$ 0.11 & 64.01 $\pm$ 6.38 
\\
& w/o prior
& \textbf{22.68 $\pm$ 1.62} & 29.91 $\pm$ 2.35 & \textbf{3.23 $\pm$ 0.32} & \textbf{97.57 $\pm$ 1.60} 
& \textbf{8.13 $\pm$ 1.63} & 15.44 $\pm$ 2.59 & 1.39 $\pm$ 0.18 & \textbf{67.97 $\pm$ 5.98} 
\\
\thline
\end{tabular}
\vspace{-0.3em}
\end{table*}

\begin{table*}
\centering
\caption{
    Results in cases (iii) and (iv): 4 microphones per array, with $T_{60} \approx$ 0 and 200~ms, respectively.
}
\label{table:cases_3_4}
\vspace{-1em}
\begin{tabular}{clcccccccc}
\thline
\multicolumn{2}{c}{\multirow{2}{*}[-1ex]{Method}} & \multicolumn{4}{c}{\multirow{1}{*}[-0.5ex]{Case~(iii)}} & \multicolumn{4}{c}{\multirow{1}{*}[-0.5ex]{Case~(iv)}}
\\
\cmidrule(lr){3-6}\cmidrule(lr){7-10}
\multicolumn{2}{c}{}    & SDR {[}dB{]} & SIR {[}dB{]} & PESQ & STOI {[}\%{]} & SDR {[}dB{]} & SIR {[}dB{]} & PESQ & STOI {[}\%{]}
\\
\midrule
\multicolumn{2}{c}{Input} 
& 0.13 $\pm$ 0.12 & 0.13 $\pm$ 0.12 & 1.10 $\pm$ 0.04 & 53.10 $\pm$ 6.21 
& 0.13 $\pm$ 0.13 & 0.13 $\pm$ 0.13 & 1.12 $\pm$ 0.05 & 51.30 $\pm$ 6.24 
\\
\multicolumn{2}{c}{Conv. (NMF)~\cite{2022KagimotoIROS}} 
& 15.91 $\pm$ 0.70 & 19.25 $\pm$ 0.82 & 1.97 $\pm$ 0.25 & 90.00 $\pm$ 3.06 
& 7.85 $\pm$ 1.34 & 14.18 $\pm$ 1.57 & 1.37 $\pm$ 0.15 & 64.59 $\pm$ 5.45 
\\
\multicolumn{2}{c}{Conv. (NTF)~\cite{2024AyanoEUSIPCO}} 
& 16.72 $\pm$ 2.00 & 29.97 $\pm$ 2.45 & 3.05 $\pm$ 0.30 & 94.61 $\pm$ 2.46 
& 7.94 $\pm$ 1.49 & 21.42 $\pm$ 1.94 & \textbf{1.53 $\pm$ 0.16} & 67.82 $\pm$ 5.69 
\\
\hline
\multirow{5}{*}{Prop.} 
& $\ShapeParam = 1$
& 15.88 $\pm$ 2.21 & 29.66 $\pm$ 2.70 & 2.23 $\pm$ 0.35 & 91.81 $\pm$ 2.50 
& 6.86 $\pm$ 1.75 & \textbf{22.68 $\pm$ 3.56} & 1.20 $\pm$ 0.08 & 63.81 $\pm$ 5.78 
\\
& $\ShapeParam = 10^{-1}$
& 18.46 $\pm$ 2.11 & 31.12 $\pm$ 2.52 & 2.63 $\pm$ 0.33 & 94.49 $\pm$ 1.87 
& 8.23 $\pm$ 1.74 & 21.82 $\pm$ 3.42 & 1.32 $\pm$ 0.12 & 68.36 $\pm$ 5.55 
\\
& $\ShapeParam = 10^{-2}$
& 18.74 $\pm$ 2.08 & \textbf{31.23 $\pm$ 2.47} & 2.68 $\pm$ 0.32 & 94.75 $\pm$ 1.80 
& 8.36 $\pm$ 1.74 & 21.64 $\pm$ 3.39 & 1.34 $\pm$ 0.12 & 68.74 $\pm$ 5.54 
\\
& $\ShapeParam = 10^{-3}$
& 18.77 $\pm$ 2.08 & \textbf{31.23 $\pm$ 2.46} & 2.68 $\pm$ 0.32 & 94.78 $\pm$ 1.79 
& 8.37 $\pm$ 1.74 & 21.62 $\pm$ 3.39 & 1.34 $\pm$ 0.12 & 68.78 $\pm$ 5.54 
\\
& w/o prior
& \textbf{22.85 $\pm$ 1.61} & 30.39 $\pm$ 2.19 & \textbf{3.27 $\pm$ 0.29} & \textbf{97.66 $\pm$ 1.43} 
& \textbf{8.97 $\pm$ 1.50} & 16.60 $\pm$ 2.24 & 1.47 $\pm$ 0.20 & \textbf{70.95 $\pm$ 5.11} 
\\
\thline
\end{tabular}
\vspace{-0.5em}
\end{table*}

\vspace{-0.1em}
\subsection{Discussion for computational complexity}
\label{ssec:discussion_computational_complexity}

Next, we evaluate the computational costs of the proposed and conventional methods.
If we naively count the number of arithmetic operations in the update rules per iteration, the computational complexity can be approximated as $O(\NumArrays^{2}\NumMics\NumFreqbins\NumFrames)$ for Prop., $O(\NumBases_{\mathrm{NMF}}\NumArrays\NumFreqbins\NumFrames)$ for Conv. (NMF), and $O(\NumBases_{\mathrm{NTF}}\NumArrays\NumFreqbins\NumFrames)$ for Conv. (NTF), where $\NumBases_{\mathrm{NMF}}$ and $\NumBases_{\mathrm{NTF}}$ denote the numbers of bases in the PF stages of Conv. (NMF) and Conv. (NTF), respectively.
As reported in \cite{2024AyanoEUSIPCO}, sufficiently large numbers of bases are required when using NMF or NTF to achieve high performance. 
Consequently, the computational cost increases linearly with the number of bases.
By contrast, Prop. does not rely on low-rank approximations and depends only on the number of arrays and microphones, which are fixed experimental parameters and typically much smaller than the numbers of bases required by conventional methods.
These results indicate that the proposed method can achieve superior performance with a substantially lower computational cost.

\section{Conclusion}
\label{sec:conclusion}

In this paper, we propose a new spotforming method for efficient post-filter estimation by utilizing cross-array non-target components.
The proposed method exploits the observation that non-target components aligned with the target speech direction in one array can be spatially separated when viewed from other arrays.
The experimental results demonstrated that the proposed method outperforms conventional methods and confirmed the effectiveness of the prior distribution.

\bibliographystyle{IEEEtrans.bst}
\bibliography{reference}

\end{document}